# Adsorption of polyelectrolytes on silica and gold surfaces

M. A. Balderas Altamirano[1], R. Camacho[2], and E. Pérez[1,2] and A. Gama Goicochea[3†]

[1]Instituto de Física, Universidad Autónoma de San Luis Potosí, Av. Álvaro Obregón 64, 78000 San Luis Potosí, S.L.P., Mexico

[2] Doctorado Institucional en Ingeniería y Ciencia de Materiales (DICIM-UASLP), Universidad Autónoma de San Luis Potosí, Av. Álvaro Obregón 64, 78000 San Luis Potosí, S.L.P., Mexico

[3] División de Ingeniería Química y Bioquímica, Tecnológico de Estudios Superiores de Ecatepec, Av. Tecnológico s/n, 55210, Estado de México, Mexico.

## ABSTRACT

The results of a study that helps understand the mechanisms of adsorption of polyelectrolytes on particles, using numerical simulation methods, specifically the one known as dissipative particle dynamics are reported here. The adsorption of cationic polyelectrolytes of two different polymerization degrees interacting with two types of surfaces, one made of gold and the other of silica is predicted and compared. We find that a more negatively charged wall does not necessarily adsorb more cationic polyelectrolytes because the electrostatic repulsion between the wall and the polyelectrolytes is stronger. Additionally, intra chain repulsion plays an important role, because the largest polyelectrolyte chains have larger excluded volume than the shorter ones. In regard to the adsorption dependence on the polyelectrolyte polymerization degree we find that the excluded volume drives the adsorption throughout the intra–chain electrostatic repulsion, because the $SiO_2$ surface is strongly negative. These results are expected to be useful for several nanotechnological applications of current interest, such as in gene therapy and in the improvement of drug delivering mechanisms.

[†]Corresponding author. Electronic mail: agama@alumni.stanford.edu

# I. Introduction

Colloidal stability is an important issue for science, technology and also for several industrial applications. This phenomenon is relevant on medical applications such as the formation of micro and nano emulsions [1], in the transport of drugs [2], and in products such as paints [3], food [4], etc. Many of these mixtures are either unstable or stable over a short period of time. The addition of polymers helps maintain the stability of the mixture for longer periods of time. Furthermore, there is a growing interest in the application of colloidal mixtures with gold nano particles [5] and silicon oxide nanoparticles [6]. Gold nanoparticles are of interest because they are inert (that is, its low reactivity in almost any environment in general), and for their antibacterial properties. On the other hand, $SiO_2$ nanoparticles are important for their applications as a means to carry drugs to increase their bioavailability or to reduce the medical dose prescribed to a patient [7, 8]. Another important application is in the production of nano shells, which arises from the discovery of materials such as SBA [9] and MCM [10]. Nano shell materials reduce the difficulties and the cost of production of nano particles that help transport materials to specific sites in the body, which have high surface area and can be functionalized (i.e., added compounds on the surface to promote reactions) with ad hoc molecules to increase the effectiveness of medical treatment, or promote specific reactions. Materials such as gold and silicon oxide nanoparticles have the characteristic of having a negative surface charge.

The procedure through which a nanoparticle gets introduced into the cell is by endocytosis, which consists of covering the particle with a positive substance like a positively charged polymer or polyelectrolytes like DNA, proteins or peptides. A positive coating helps nano particles approach the cell membrane and be embedded in the cell. This application may be feasible for gene therapy, where a molecule or encapsulated gene is introduced in the cell and released inside; the gene is mixed with the DNA of the cell and produces a reaction that strengthens the DNA. A material typically used to coat nano particles is polyethylimine (PEI), which is a polymer made of ethylamine monomers. PEI has positive charges, is not toxic and is also bioavailable. PEI is frequently replaced by polyethylenoxide (PEO), chitosan, polypeptides, etc.

One important phenomenon present in all these applications is adsorption, which arises from the interactions between the particles and, in this case, polyelectrolytes. When a solid surface interacts or adsorbs polyelectrolytes, these cover and change the surface of the solid particle. Sometimes the polyelectrolyte needs to compensate the surface charge; this mechanism is called charge reversal or overcompensation [11]. The adsorption process of polyelectrolytes on oppositely charged surfaces has been studied for a long time, and there are plenty of works on this area, including experimental, theoretical [12-17] and computer simulations [18-19] studies. One important theoretical work is Manning's theory of counterion condensation [20], which explains why the total charge on a strongly charged PEI chain is lower than the formal value expected from the chemical structure of PEI and the ionic strength. Some other works have been published on this matter, and the reader is referred to [21]. Most of these works consider the adsorption over an electrically neutral surface, however real surfaces usually have superficial charge densities. In this work we add a charge distribution over the surface and a correction of the Ewald sums to fully take into account the confinement [22].

Our model considers gold and silica particles that are much larger than the polyelectrolyte chains, which are dispersed in an aqueous solvent that contains the chains and the counter ions. Because of the difference in size between the particles and the rest of the molecules, we assume they can be modeled as planar walls. Hence, two neighboring particles would form two parallel walls confining the solvent, polyelectrolytes and counter ions. Lastly, because the oxide particles are much more massive than the rest of the constituents of the fluid, their dynamics is much slower than that of the fluid molecules, which allows us to fix the planar walls in space and focus all computational efforts in solving the dynamics of the fluid. We perform Monte Carlo simulations in the Grand Canonical ensemble to allow for the exchange of solvent molecules between the region of fluid confined by the oxide particles and the surrounding fluid. It is assumed that the system has reached full thermodynamic equilibrium and that all polyelectrolytes and counterions have been adsorbed or have formed associations in the fluid without leaving the confined region, therefore we fix the chemical potential of the solvent molecules only.

Here we model a solid wall (nano particle, or colloid) interacting with polyelectrolytes of two different lengths (P10 or P100) and include the surface charge density appropriate for Au or $SiO_2$ surface, as modeled by Alarcón et al. [22], using coarse grained Monte Carlo simulations. To model systems with so disparate sizes such as colloids, solvent molecules and polyelectrolyte chains one needs to go beyond atomistic molecular dynamics (MD) schemes, which are very time – consuming [23]. There are alternative methods, such as dissipative particle dynamics (DPD) [24], which follows similar rules to integrate the equation of motion as MD, for beads which are thought of as a grouping of particles, thereby defining the coarse grained reach of the simulation. In Section II we report the methodology and details of our simulations; Section III is devoted to the presentation of the results and their discussion. The conclusions are given in Section IV.

**II. Models, Methods and Simulation Details**

MD and DPD both solve Newton's second law for a system of particles as time evolves; the difference between them comes with the model interactions. In DPD one considers a conservative force ($F_i^C$) acting between pairs of particles, which decays linearly and is of short range. There are two extra forces in the model; the Dissipative ($F_{ij}^D$) and ($F_{ij}^R$) Random forces, which constitute a built-in thermostat. The total force acting on any particle $i$ is given by:

$$\boldsymbol{F}_i = \boldsymbol{F}_i^C + \sum_{i \neq j}^{N} [\boldsymbol{F}_{ij}^D + \boldsymbol{F}_{ij}^R] \qquad (1)$$

.

The dissipative force $F_{ij}^D$ (3) and random force $F_{ij}^R$ (4), are defined as follows:

$$\boldsymbol{F}_{ij}^D = -\gamma(1 - r_{ij}/r_c)^2 [\hat{\boldsymbol{e}}_{ij} \cdot \boldsymbol{v}_{ij}]\hat{\boldsymbol{e}}_{ij} \qquad (2)$$

$$\boldsymbol{F}_{ij}^R = \sigma(1 - rij/r_c)\xi_{ij}\hat{\boldsymbol{e}}_{ij} \qquad (3)$$

where $r_{ij} = r_i - r_j$ is the relative position vector, $\hat{\boldsymbol{e}}_{ij}$ is the unit vector in the direction of $r_{ij}$ and $\boldsymbol{v}_{ij} = v_i - v_j$ is the relative velocity, $r_i$ and $v_i$.are the position and velocity of particle i respectively. The random variable $\xi_{ij}$ is generated between 0 and 1 with a Gaussian distribution of unit variance. The constants $\sigma$ and $\gamma$ are the strength of the random and dissipative forces, respectively; $r_c$ is a cut off distance. All these forces are zero for $r_{ij} > r_c$. All beads are of the same size, with radius $r_c$, which is set to equal 1. The dissipative and random forces are coupled through the fluctuation dissipation theorem which leads to the next expression [24]:

$$k_B T = \frac{\sigma^2}{2\gamma} , \qquad (4)$$

where $k_B$ is the Boltzmann constant and $T$ the absolute temperature. The system is composed of linear chains (P10 or P100) made up of DPD beads confined, in a simulation cell with two walls on the $z$ – axis; these walls have a specific surface density charge. We have three kinds of DPD beads, which interact with the following short range force:

$$\boldsymbol{F}_{ij}^{ExcVol} = \begin{cases} a_{ij}(1 - r_{ij})\hat{\boldsymbol{r}}_{ij} & r_{ij} \leq r_c \\ 0 & r_{ij} > r_c \end{cases} \qquad (5)$$

$F_{ij}^{ExcVol}$ is the force between the beads of the system, that models the excluded volume interaction.

We use the Kremer-Grest model for linear polymers [25] where beads are bonded through a freely rotating harmonic spring:

$$\boldsymbol{F}_{ij}^{spring} = -k_0(r_{ij} - r_0)\hat{\boldsymbol{e}}_{ij}, \qquad (6)$$

where $k_O$, is the spring constant and $r_O$ is the equilibrium distance; for our calculations we use $k_O = 100$ and $r_O = 0.7$ [26].

The polymers modeled here have charge, therefore we apply Ewald sums [27], as is customary in numerical simulations to fully take into account the long – range nature of the electrostatic interactions. Under this approach, the total electrostatic interaction is divided into two parts: one in real space $\left(F_{ij}^{E,R}\right)$, plus one in Fourier space $\left(F_i^{E,K}\right)$. For distributions of charge, rather than point charges as is necessary for DPD beads, the expressions for those two forces read as follows [28]:

$$\boldsymbol{F}_{ij}^{E,R} = \frac{\Lambda}{4\pi} Z_i Z_j \left[\frac{2\pi}{V}\exp(-\alpha^2 r_{ij}^2) + erfc(\alpha r_{ij})\right] \quad (7)$$

$$\times \left\{1 - e^{(-\beta r_{ij})}\left[1 + 2\beta r_{ij}(1 + \beta r_{ij})\right]\right\}\frac{\mathbf{r}_{ij}}{r_{ij}^3}$$

$$\boldsymbol{F}_i^{E,K} = \frac{\Lambda}{4\pi} Z_i \left\{\frac{2\pi}{V}\sum_{k\neq 0}^{\infty} Q(k)\mathbf{k} \times Im\left[e^{(-i\mathbf{k}\cdot\mathbf{r}_i)}S(\mathbf{k})\right]\right\}, \quad (8)$$

where $\Lambda = \frac{e^2}{k_B T \varepsilon_o \varepsilon_r r_C}$, with $e$ being the electron's charge, $\varepsilon_o$ is the permittivity of vacuum, $\varepsilon_r$ = 78.3 is water's relative dielectric permittivity at room temperature; $\beta = \frac{r_C}{\lambda}$, and $Z_i$ is the valence of the charge distribution; $\boldsymbol{k} = 2\pi\left(\frac{k_x}{L_x}, \frac{k_y}{L_y}, \frac{k_z}{L_z}\right)$ is the reciprocal vector of magnitude $k$, such that $k_x$, $k_y$, and $k_z$ are integers, $Q(\boldsymbol{k}) = exp\left(-\frac{k^2}{4\alpha^2}\right)$ and $S(\boldsymbol{k})$ is the so called structure factor [27]. The symbol $\alpha$ refers to the parameter that determines the contribution of the sum in real space, see eq. 7. The term $erfc(x)$ is the complementary error function; $V = L_x L_y L_z$ is the volume of the simulation box. Here, $Im$ denotes the imaginary part of the complex number.

To represent the oxide surface of the nanoparticles, we used a planar wall, defined by the following effective force $F_i^W$, which is the force between a wall and a bead in the fluid system, as first proposed in [3]:

$$F_i^W(z) = a_w \left[1 - \frac{z}{R_c}\right]. \qquad (9)$$

In the equation above, $a_w$ is the strength of the DPD conservative interaction between a particle in the fluid and a surface particle. Equation (9) becomes identically zero for distances $z > R_C$. Equations 7 and 8 are used for fluids with three – dimensional periodicity. To simulate confined systems using Ewald sums one needs to remove the net dipole moment of the simulation box, adding an extra force to each particle in the fluid [29]:

$$F_i^Z = -\frac{\Lambda q_i}{V} M_Z, \qquad (10)$$

where $M_z$ is the net dipole moment of the simulation cell, which is given by

$$M_Z = \sum_{i=1}^{N} q_i z_i. \qquad (11)$$

Such dipole moment must be removed out of the simulation cell for each charge $q_i$, to avoid artifacts introduced by the confinement [22]. An additional force used in this work is $F_i^{EW}$, which models the force between the charges on the surface of the nanoparticles and the charge distributions in the DPD fluid particles, as first proposed in [22]:

$$F_i^{EW}(z) = \Lambda k Z_i e^{-\frac{2\beta}{\lambda}} \left[\frac{1}{z} - 2\beta \ln(z)\right], \qquad (12)$$

where $\Lambda$ and $\beta$ were defined before (see text following eqn. 8) and

$$k = \frac{a_z^2 Z_{OX}}{\pi \lambda^3}. \qquad (13)$$

Here, $Z_i$ is the valence of the fluid particle $i$, $Z_{ox}$= -1.3 for gold and $Z_{ox}$= -11.3 for silica particles and $a_z = 1.27 \, \text{Å} = 0.97 r_c$. This force leads to a charge density that is uniformly distributed on the effective surfaces.

The total conservative force is:

$$F_i^C = \sum_{i \neq j}^{N} [F_{ij}^{ExcVol} + F_{ij}^{spring} + F_{ij}^{E,R}] + F_i^{E,K} + F_i^W + F_i^Z + F_i^{E,W}$$

(14)

where $F_i^C$ is the total conservative force acting on particle $i$, see eqn. 1.

We use the Gran Canonical ($\mu VT$) ensemble to ensure the fluid modeled is in chemical, mechanical and thermal equilibrium [35]. In this ensemble, to keep the chemical potential ($\mu$) constant, it is necessary to add or delete water beads; the Monte Carlo (MC) algorithm was used to carry the adsorption simulations [27]:

$$P_{insertion} = min\left[1, \frac{\langle Z(z) \rangle V}{N+1} exp\left(-\frac{\Delta U^{test}}{k_B T}\right)\right] \qquad (15)$$

$$P_{deletion} = min\left[1, \frac{N}{\langle Z(z) \rangle V} exp\left(-\frac{\Delta U^{test}}{k_B T}\right)\right] \qquad (16)$$

(here *min* [a, b] indicates that the minimum between a and b is to be chosen ), where $\Delta U^{test}$ is the total conservative interaction difference between the added or removed bead, and the $N$ or $N$ -1 remaining beads, including the conservative interaction with the surface.

$\langle Z(z) \rangle = e^{-\mu/kT}$ is the so called activity, which is determined by the chemical potential [35].

We carried out simulations for four types of systems: one is the adsorption of a linear polyelectrolyte made of ten beads, called P10. The other is made up of one hundred beads (P100); both are confined by two different types of surfaces, gold and silica oxide. For each system we performed simulations with 100, 200, 300, 400, 500, 600, 700, 800 beads that make up the P10 or P100 polyelectrolytes. In each simulation the number of water beads fluctuates during the simulation process, to keep the chemical potential constant. The simulations results were obtained after at least 100 blocks of $10^4$ MC configurations each one, with the first 30 blocks used to equilibrate the system. In each given block made up of $10^4$ MC configurations, the percentage of successful MC moves was around $(35 \pm 2)$. Dimensionless unit are used throughout this work. The size of the simulation box was set at $L_x=L_y=20$ and $L_z=10$, periodic boundary conditions were used in the $x$, $y$ directions. The chemical potential $\mu*$ was fixed at 37.7, which leads to an average density $\langle \rho^* \cong 3 \rangle$ [30]. The time step was $\delta t = 0.03$ for the DPD part, where the DPD bead were moved using the Velocity Verlet algorithm [27]. The parameters that define the dissipative and random forces are $\gamma = 4.5$ and $\sigma = 3.0$, so that $k_B T = 1$. The conservative force intensities were chosen as $a_{ij} = 78.3$, when $i = j$ and $a_{ij} = 79.3$ when $i \neq j$, where $i$ and $j$ represent types of fluids particles (solvent, polyelectrolyte or counterion), with the exception of the polyelectrolyte-counterion interaction, which is set at $a_{ij} = 78.3$. Additionally, $a_w = 120.0$, for the solvent-wall, non-electrostatic interaction, and $a_w = 60.0$ for all the other molecular species wall interaction [27]. This choice of wall – bead interactions ensures the adsorption of the polyelectrolytes and counterions on the surfaces. The solvent and the counterions are modeled as simple monomeric particles. All beads in the polyelectrolytes are freely joined by harmonic springs. The parameters of the Ewald sum were $\alpha = 0.11 \text{Å}^{-1}$ and the maximum vector $k^{max} = (6,6,6)$. The values of $\beta = 0.929$, $\lambda = 6.95 \text{Å}$ and $\Lambda = 13.87$ were used as in ref [22]. Each bead on the polyelectrolytes has charge of $0.5e$ and $-0.5e$ for each counterion. We take $q_{Ox} = -1.15e$ for Au surfaces and $-11.5e$ for SiO$_2$ surfaces; with the above choice of parameters, the surface charge density on the walls is $\sigma_{OX} = 60 \, mCm^{-2}$ for Au and $\sigma_{OX} = 600 \, mCm^{-2}$ for SiO$_2$. Length was normalized with $r_c = 6.46 \text{Å}$, which is the value that corresponds to a coarse graining degree of 3 water molecules in one DPD bead [30]. This length is also the maximum range of the interparticle forces (see eqn 2, 3, 5) and the non-electrostatic wall force (see eqn 9). Full details of the method have been published elsewhere [22].

Adsorption isotherms were calculated using the density profile of the polyelectrolytes; the same procedure was carried out for the adsorption of counterions on the oxide surfaces.

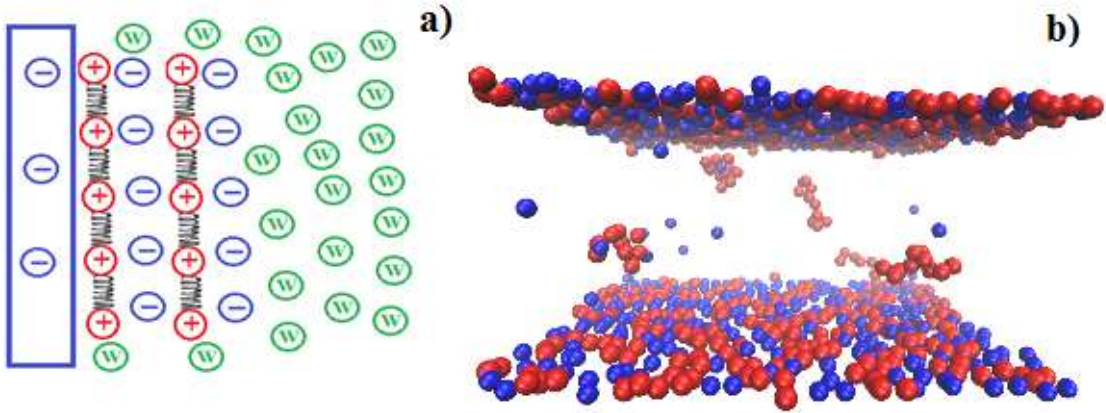

*Figure 1. (Color online) Polyelectrolyte adsorption model. Fig. 1a Polyelectrolytes made up of beads bound with springs (red beads); counterions are shown in blue, and in green water molecules, (W). Fig 1b) shows a snapshot of a typical simulation, for polyelectrolytes with 300 beads, or 30 chains of P10, on the Au surface. The solvent particles have been removed for simplicity.*

*Figure 1a* shows a schematic depiction of the phenomenon simulated in this work. The rectangle on the left represents the gold or silicon oxide nanoparticles. The circles with a + sign, are the polyelectrolyte beads, connected by springs. Circles with negative signs represent the counter ions, and circles with *W* are water molecules. *Fig. 1b* is a snapshot of a simulation at a concentration of 30 chains of the P10 type adsorbed on a gold nanoparticle; notice that the polyelectrolytes and counterions adsorb on the surfaces uniformly, while a few polyelectrolyte chains and counterions remain in the bulk phase. The solvent particles were removed for simplicity.

**III. Results and Discussion**

We start by presenting the density profiles, the adsorption isotherms and finally we discuss the electrostatic potential profile.

*III.1 Density profiles*.

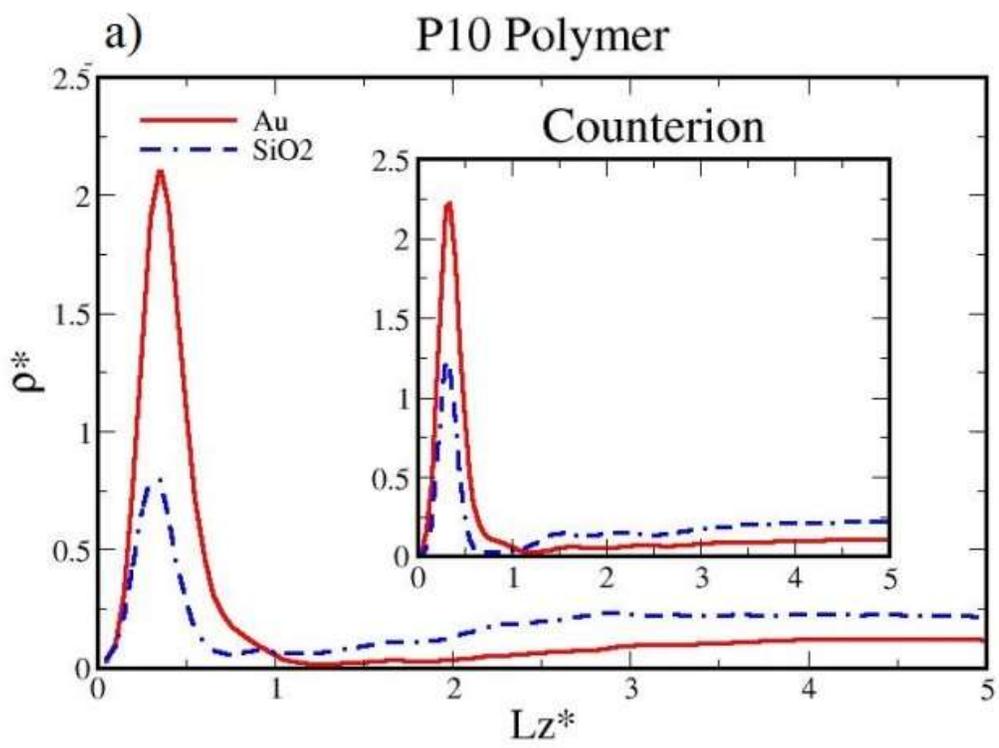

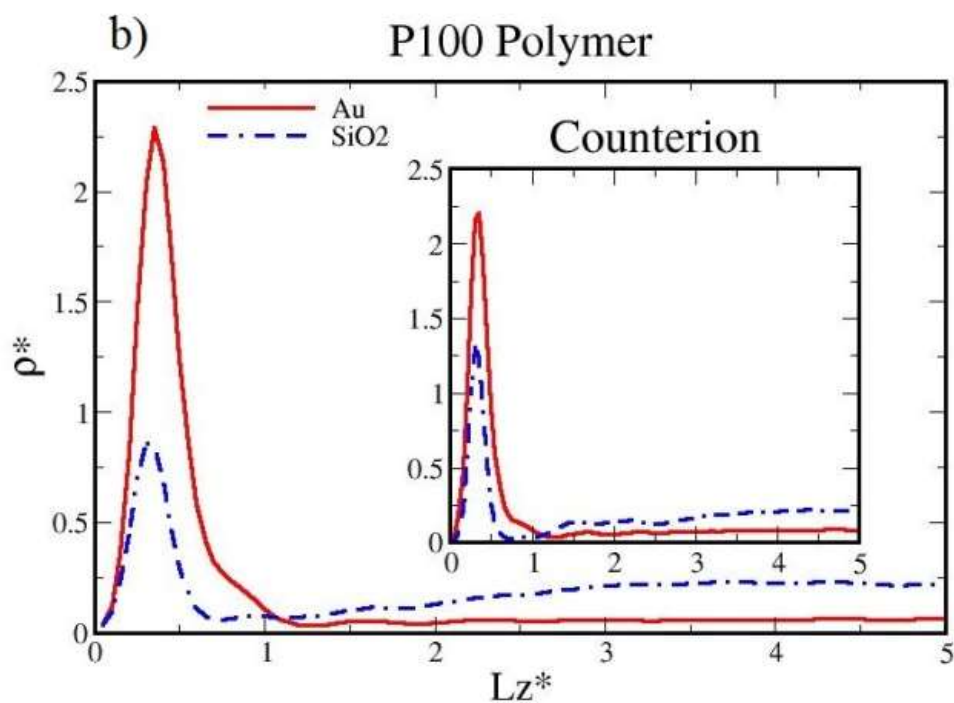

*Figure 2 Density profiles at a concentration of 800 polyelectrolyte beads on Au and SiO2. Fig 2a corresponds to the density profile of P10, and Fig2b to the P100 polymer. The insets show the counterion density profiles of each simulation. For both the P10 and P100 polyelectrolytes, the solid lines correspond to adsorption on Au, be it chains or counterions, and broken lines represent adsorption on the SiO$_2$ surface. Fig2a corresponds to 80 P10 chains and Fig 2b to 8 P100 chains. The profiles shown here are averaged over both sides of the box.*

*Fig. 2* shows the density profile of the P10 (*fig 2a*), and P100 (*fig 2b*) polyelectrolytes at a concentration of 800 beads, namely 80 chains of P10 or 8 of P100. The solid lines represent the adsorption on gold surface and broken lines the adsorption on SiO$_2$ surface. Each graph has an inset, which corresponds to the counterions density profile. These graphs were also obtained dividing the *z* axis in two and averaging the density profile. Both profiles shown in *Fig 2a* and *2b*, clearly show that the adsorption of the polyelectrolytes (P10 and P100), is greater on the gold surface than on the silicon oxide surface, although the latter is more negatively charged. Another important observation is the similarly adsorbed amount of counterions and polyelectrolytes on the gold wall, while in the case of the $SiO_2$ surface the counterion adsorption is slightly larger than the polyelectrolyte's. This is the result of the relatively small charge on the surface of the gold particles, therefore electrostatic attraction (for polyelectrolytes) or repulsion (for counterions) is not the leading adsorption mechanism. Instead, the van der Waals – type of interaction represented by eqn 9 between the surface and the polyelectrolytes or counterions is dominant, and since the magnitude of such interactions is the same for both types of molecules (polyelectrolytes and counterions), their adsorption on the Au surface is very similar; see also the snapshot in Fig. 1b. A different adsorption mechanism takes place when the surface charge is relatively large, as is the case for the $SiO_2$ particle, because electrostatic attraction would lead to more polyelectrolyte adsorption, but since the polyelectrolytes carry charge themselves each of their beads creates an ionic radius on the surface, which reduces adsorption. Counterion adsorption on $SiO_2$ is only slightly larger than the polyelectrolytes', simply because they are monomeric and therefore, more easily accommodated on the surface. Lastly, we comment on the effect of the polymerization degree on the adsorption. As Fig.2b shows, the P100 polyelectrolyte adsorption on Au is about ten percent larger than that of the P10 polyelectrolyte on the same surface, and this is clearly due to the larger number of topological connections of the P100 polyelectrolyte. Because of this increased adsorption, less P100 polyelectrolyte remains in the bulk, compared with the P10 case, as expected.

*III-2 Monolayer of polyelectrolytes and counterions.*

In this section we present adsorption isotherms for all the cases studied, and their best fits to the Langmuir adsorption model.

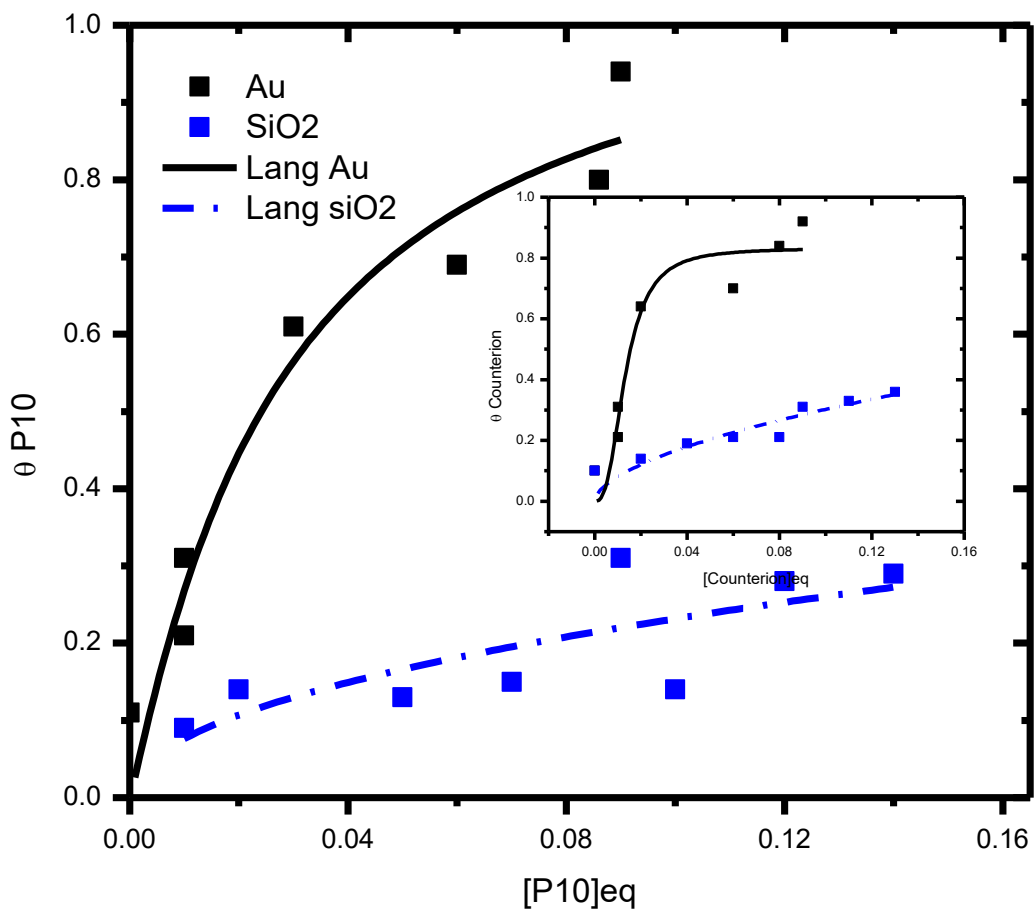

*Figure 3. Adsorption isotherms of P10 polyelectrolytes on Au and SiO2 surfaces. Black squares show the P10 polyelectrolytes or counterions (inset) simulation data on Au surfaces; blue squares (simulation data) show the adsorption of the P10 polyelectrolytes on SiO2 (see the counterions behavior in the inset). Continuous and broken lines represent the best fits using equation 17.*

*Fig 3* presents the adsorption results for the polyelectrolyte P10 on both types of surfaces. To construct the isotherms (*Fig 3*) we used the density profiles of *Fig* 2 and calculated the concentration of polyelectrolytes or counterions on the surface. The solid and broken lines represent the best fits to the Langmuir adsorption model (eq 17) for the adsorption of P10 on Au and $SiO_2$, respectively:

$$\theta = \frac{\alpha \beta c^{1-x}}{1+\beta c^{1-x}} \quad , \qquad (\mathbf{17})$$

where $\theta$ is the fraction of polyelectrolytes on the surface, $c$ is the concentration of polyelectrolytes in the bulk, $x$ is the exponent used to fit the data, $\alpha$ and $\beta$ are adjustable constants. Table I displays the parameters for eq. 17 which reproduce the correlation lines in Fig. 3. The adsorption isotherm for P10 on Au is larger than that on $SiO_2$, as the density profiles shown in Fig.2 indicate. Additionally, both isotherms predicted by our simulations and their best fits to eqn 17 show that the saturation of the $SiO_2$ surface with P10 or with counterions requires a much larger amount of those molecules than is required to saturate the Au surface, for the reasons laid out in the discussion of Fig.2.

*Table I. Correlation parameters used with equation 17.*

| Surface - Molecule | $\alpha$ | $\beta$ | $x$ |
|---|---|---|---|
| Au P10 | 3919 | 2.864e-4 | 0.4298 |
| $SiO_2$ P10 | 0.83 | 107432 | -1.68 |
| Au-Counterion | 1.097 | 46.75 | -0.08017 |
| $SiO_2$ Counterion | 2061 | 3.4013-4 | 0.5192 |

The adsorption isotherms for P100 are qualitatively and quantitatively very similar to those shown in *Fig 3* and therefore we omit them here, for brevity. For the P10 and P100 cases the adsorption of polyelectrolytes on Au is greater than that obtained at each concentration on the surface of silicon oxide. The inset in *Fig. 3* shows the adsorption isotherms of the counter ions, which is higher on Au than on the $SiO_2$ surface. These trends are found for the P100 polyelectrolytes on both types of surfaces. An important variable in these simulations is the electric charge on the surface and on the polyelectrolytes. To study the influence of the charges along the simulation box we calculated the electrostatic potential profile, which is the subject of the next section.

*III-3 Electrostatic Potential Profile*

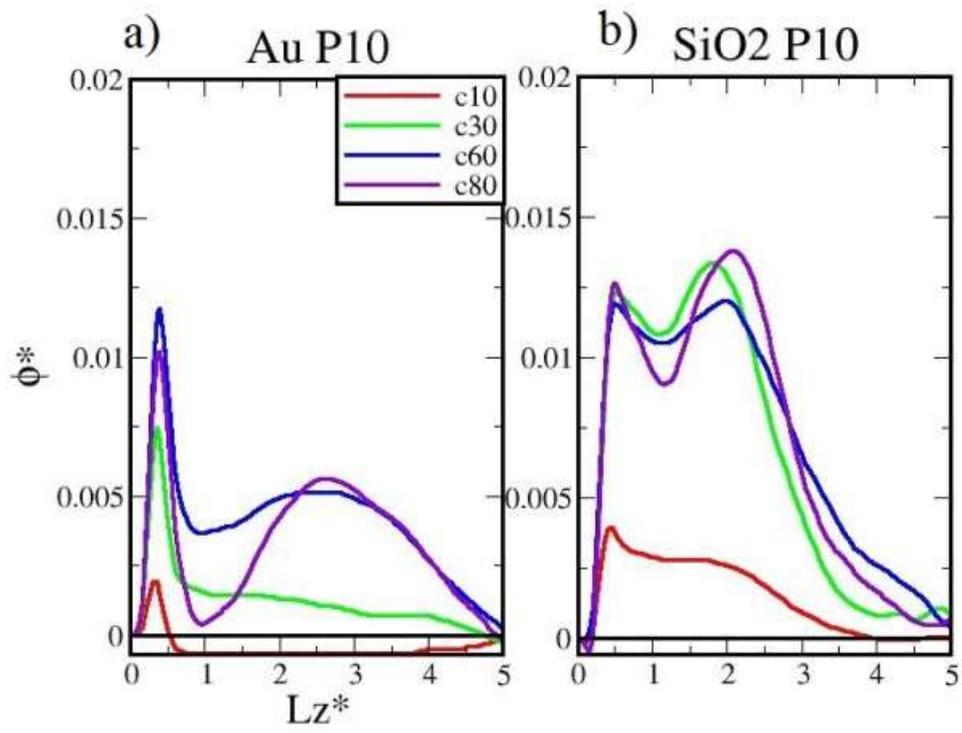

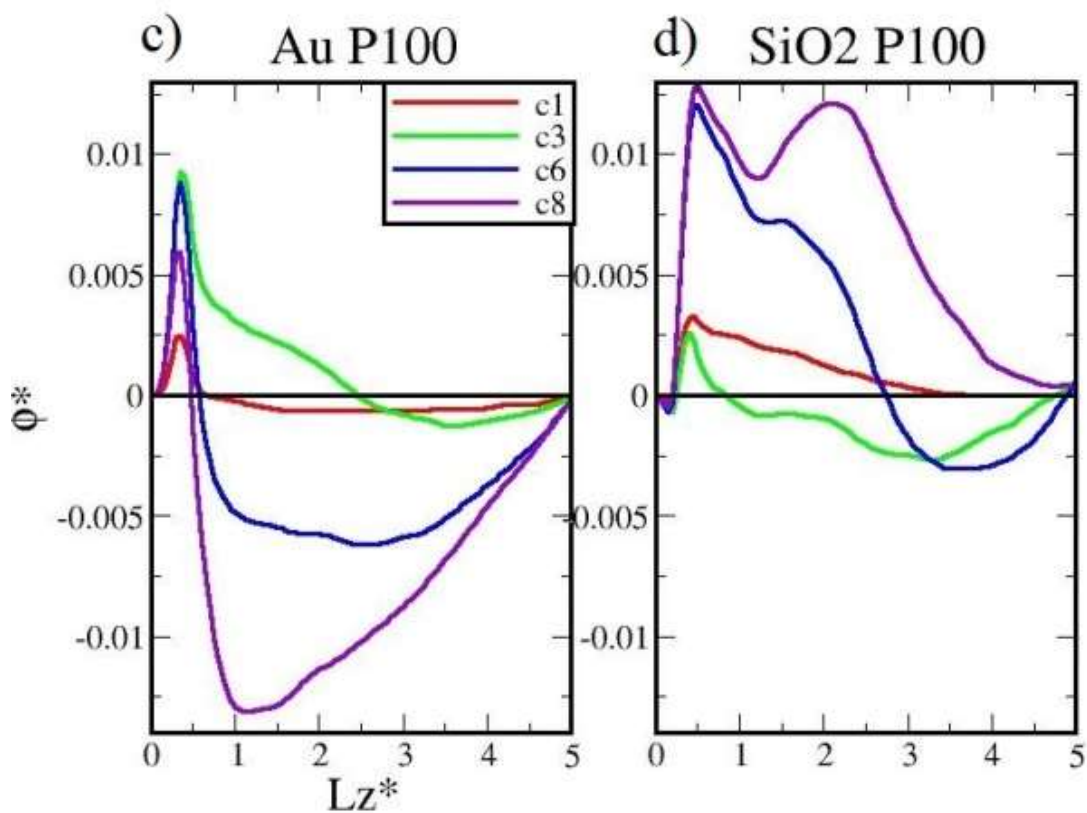

*Figure 4. (Color online) Electrostatic potential profiles at four polyelectrolyte (P10; P100) concentrations on Au and SiO$_2$ nanoparticles, as functions of the distance perpendicular to the surfaces. Fig 4a, 4b, and 4d, show increasing of the potential near the surface as the concentration of polyelectrolytes increases. Fig 4c shows charge inversion beyond the adsorbed monolayer of polyelectrolytes and counterions.*

To calculate the total electrostatic potential as a function of the distance perpendicular to the surfaces ($z$) we used the expression reported in ref [11]:

$$\varphi(z) = -\frac{e^2}{k_B T R_C \varepsilon_o \varepsilon_r} \int_z^L [\rho(z)(z-z')]dz, \quad (18)$$

where φ(z) is the total electrostatic potential and $\rho(z) = Z_+ \rho^+(z) + Z_- \rho^-(z)$, using the density profile of the $\rho^+(z)$ cation and $\rho^-(z)$ anion, respectively; $Z_+$ and $Z_-$ are the charge of the polymer or counter ion respectively. The *z* and *z'* are the positions along the *z* axis where the electrostatic potential is calculated.

The profile of the electrostatic potential is a helpful measure of Coulomb interactions in the simulation box, between the surface, the polyelectrolytes and the counterions. In this case, both types of molecules compete for the same surface and the attraction is larger for the polyelectrolytes than for the counterions, but the latter also surround the polyelectrolytes to neutralize the charge, therefore polyelectrolyte adsorption necessarily leads to counterion adsorption on the surface. The attraction for the $SiO_2$ wall is stronger for the polyelectrolytes than for the counterions. This attraction also leads to an internal repulsion between the beads that make up the polyelectrolyte. We have computed the electrostatic potential for all cases studied and the results are shown in *Fig 4*. The electrostatic potential profile corresponding to the polyelectrolytes P10 on Au is presented in *Fig. 4a)*, where there appear maxima near the surface of the nanoparticle at all polyelectrolyte concentrations. These peaks signal the formation of a well-defined monolayer of polyelectrolytes and counterions on the Au surface. Farther away from the surface a minimum and a wide second peak appear, these features are due to the formation of a second diffuse layer. In the case of the $SiO_2$ P10 system, *Fig. 4b)*, a broad peak can be seen and a double peak formation on top, which correspond to the formation of not very well defined multilayers. *Fig 4c)* shows the formation of narrow maxima in the electric potential profile close to the Au surface, corresponding to the formation of a well-defined P100 polyelectrolyte and counterion layer. As the polyelectrolyte concentration is increased those peaks increase as well. However, as the distance perpendicular to the surface increases along the *x* – axis ($L_z^*$), the electric potential decreases and becomes negative, which is the consequence of having more free (not adsorbed) counterions than polyelectrolytes away from the surface. Lastly, for the case of the polyelectrolyte P100 on the $SiO_2$ surface, shown in *Fig.4d)* there appear broad maxima close to the surface having positive electrostatic potential, which increases when the polyelectrolyte concentration increases. The electrostatic potential stays positive for wider distances from the surface because the polyelectrolyte is larger and only some of its monomers are adsorbed, leaving the rest more or less free to move in the confined fluid. A well-defined maximum close to the surface, as can be seen for the results on Au (Figs 4a and 4c), indicates there is a layer made up of polyelectrolytes and counterions adsorbed on the surface. This layer begins to disappear when the counterion and polyelectrolyte concentrations are increased in the system. From this analysis two important aspects arise that reinforce the first conclusions, one is that the polyelectrolyte layers on gold are well defined, as seen in the density profiles in Fig. 2. Additionally, the electric potential profiles of the systems with the silica surface present broad maxima, which are the consequence of the formation of not well – defined multilayers composed of counterions, polyelectrolytes and water molecules.

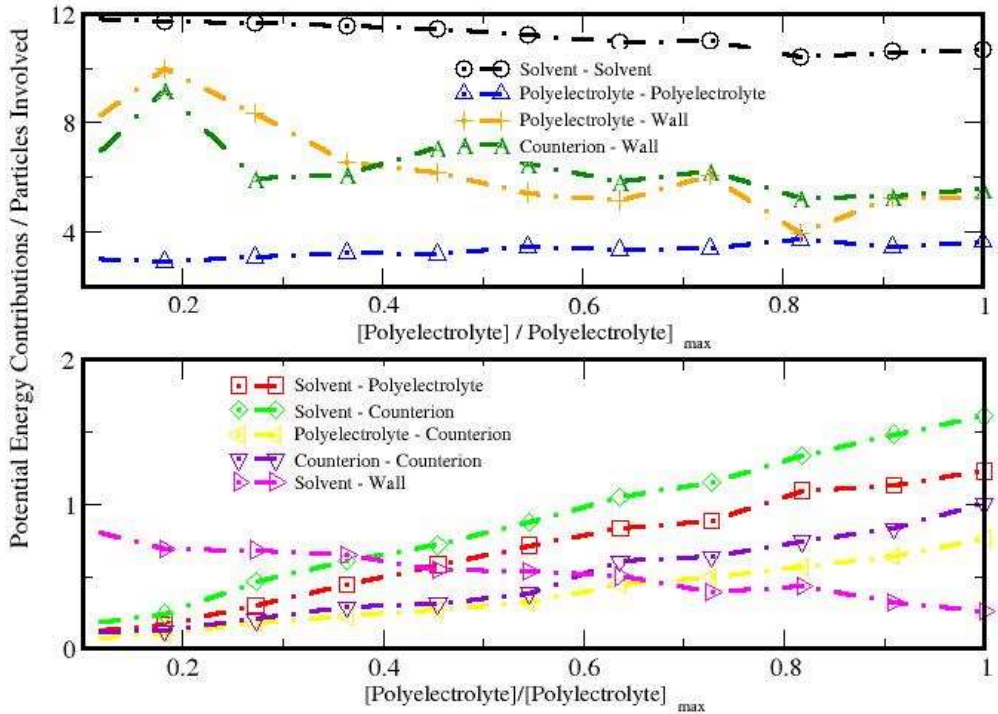

*Figure 5. (Color online) Conservative energy contributions (in reduced DPD units) of each type of particle for the P10 polyelectrolyte adsorption on the Au particles, normalized by the number of particles of each corresponding type, as functions of the polyelectrolyte concentration. The latter is normalized by the maximum concentration.*

We know analyze the contributions to the total conservative energy coming from the different pairs of particles. In Figs. 5 and 6 we show those contributions for the case of the P10 and P100 polyelectrolytes, respectively, on the Au particles. In both cases, the total conservative interaction between particles of the same type (solvent – solvent and polyelectrolyte – polyelectrolyte) remains approximately constant, except for the case of counterions, where a slight increase is observed as the concentration is increased. This is due to their increased kinetic energy because they are individual monomers, in addition to their repulsive electrostatic interaction. For both types of polyelectrolytes (P10 and P100), the interactions between all types of fluid particles and the surfaces decrease with increasing concentration because the walls get rapidly saturated (see Fig.1) and the wall interaction is of short range, see Eq. (9). The interactions of the solvent with polyelectrolytes and counterions, and polyelectrolyte – counterion interaction increase somewhat with concentration due to excluded volume interactions.

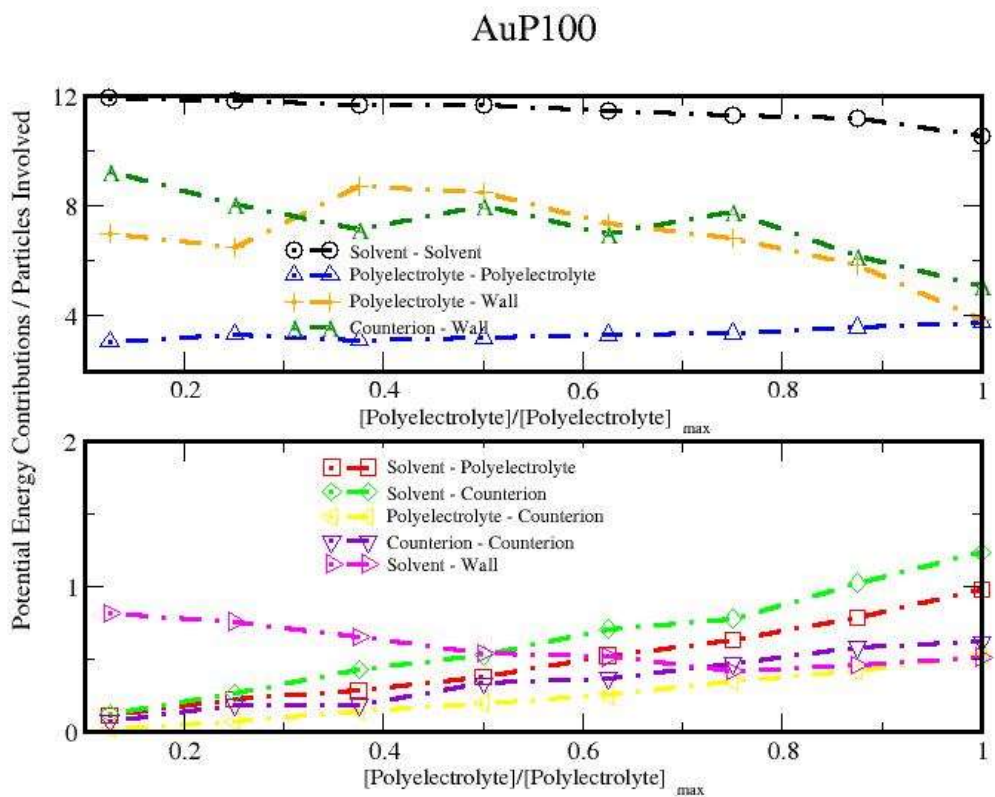

*Figure 6. (Color online) Conservative energy contributions of each type of particle for the P100 polyelectrolyte adsorption on the Au particles, normalized by the number of particles of each corresponding type. Quantities are presented in reduced DPD units.*

Figures 7 and 8 show the total conservative interactions for the case in which the $SiO_2$ walls confine the fluid containing the P10 and P100 polyelectrolytes, respectively. The trends seen in those figures are qualitatively the same as those presented in Figs. 5 and 6, except that the $SiO_2$ surface – polyelectrolyte interaction is smaller than the Au – polyelectrolyte interaction, at increasing concentration of P10. That is why adsorption of the P10 polyelectrolyte on silica is below that on gold, see Fig. 3.

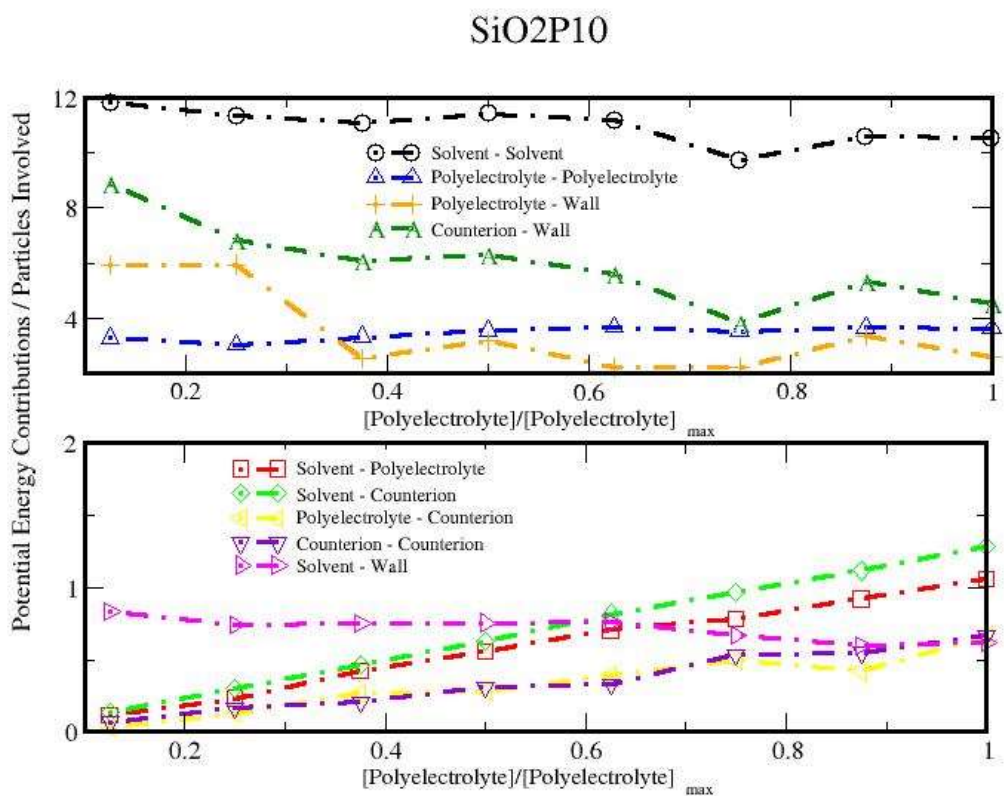

*Figure 7. (Color online) Conservative energy interactions for the P10 polyelectrolyte adsorption on the SiO$_2$ particles, in reduced DPD units.*

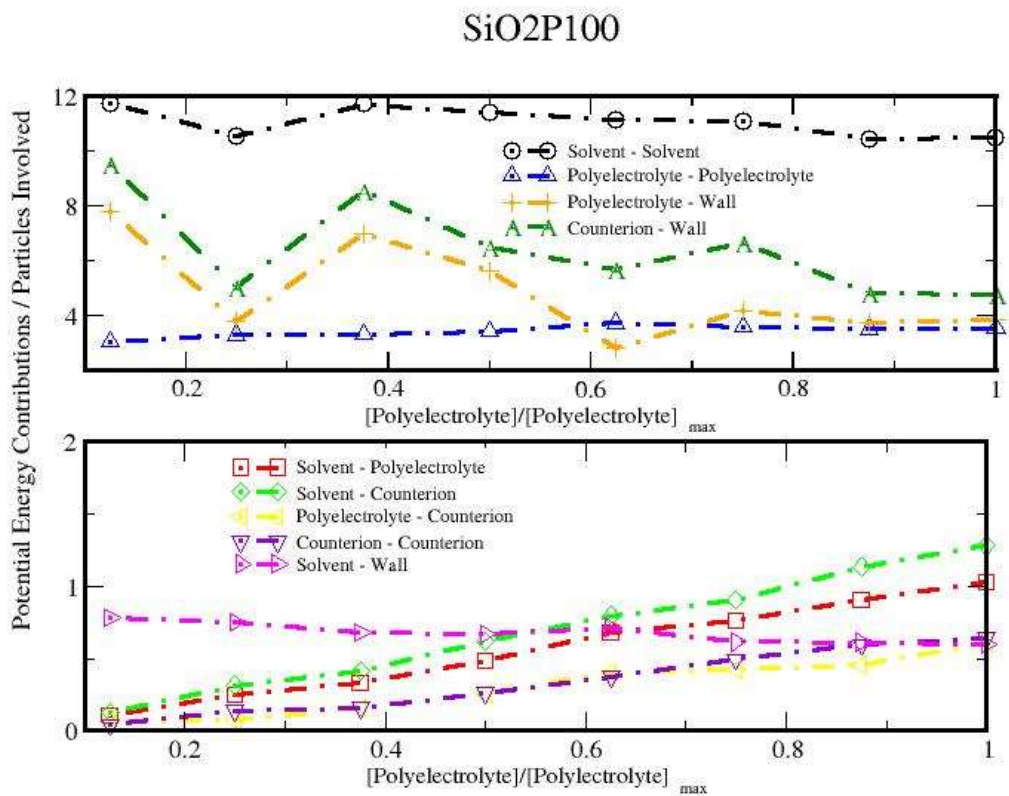

*Figure 8. (Color online) Conservative energy interactions for the P100 polyelectrolyte adsorption on the SiO$_2$ particles, in reduced DPD units.*

In Fig. 9 we present the interfacial tension between the surfaces and the fluid, calculated as the difference between the averaged components of pressure tensor normal to the walls and tangential to the them $\gamma = \langle P_N \rangle - \langle P_T \rangle$. Notably, the interfacial tension for the fluid confined by gold particles is approximately the same regardless of the polyelectrolyte length, and decreases with increasing polyelectrolyte concentration, whereas for silica walls the interfacial tension increases and is larger for the P100 polyelectrolyte.

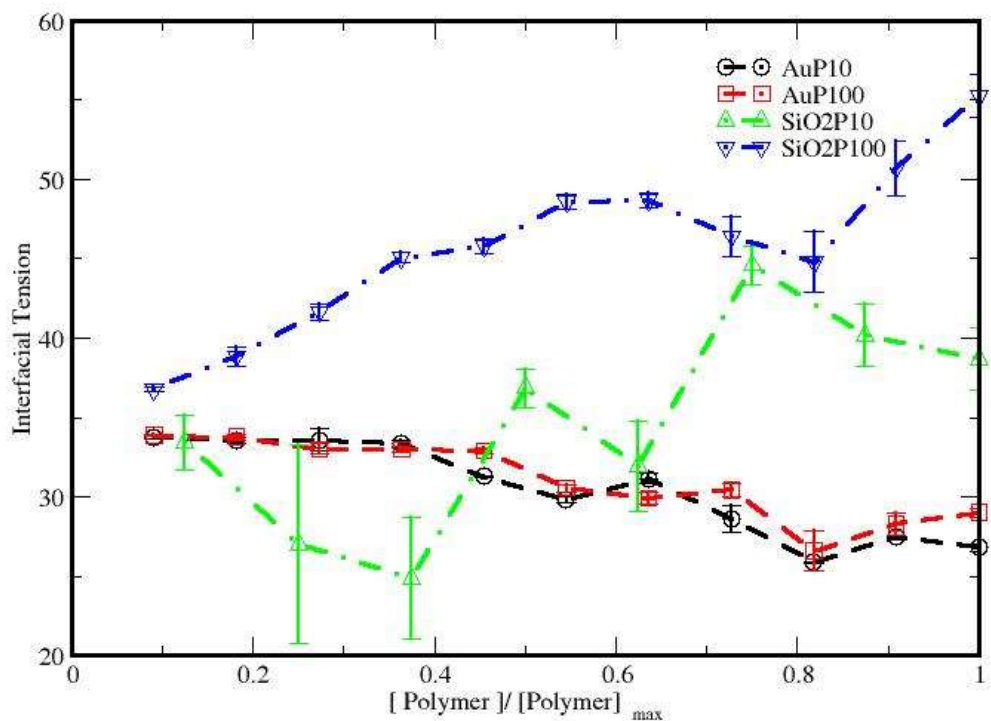

*Figure 9. (Color online) Interfacial tension between the surfaces and the fluid, in reduced DPD units, as a function of the normalized polyelectrolyte concentration.*

The interfacial tension with Au particles does not vary much with increasing P10 concentration because the polyelectrolyte is more favorably adsorbed on them than on $SiO_2$ particles, see Fig. 2, which makes the interface between the particle and the fluid more uniform. On the other hand, the adsorption of P10 on silica is weaker, which creates a less uniform interface between the wall and the fluid and this is turn translates into a larger interfacial tension, as seen in Fig. 9. The polymerization degree does not change the interfacial tension for Au particles because the surfaces are saturated at relatively low polyelectrolyte concentration, but it does come into play for $SiO_2$ surfaces, where a more irregularly covered surface (by P100) translates into larger interfacial tension. Lastly, since the interfacial tension is the change in free energy with surface area, it is to be expected that such energy cost be larger for $SiO_2$ than for Au particles because adsorption of the polyelectrolytes is smaller on silica than on gold (see Fig. 3), and that is indeed what Fig. 9 shows.

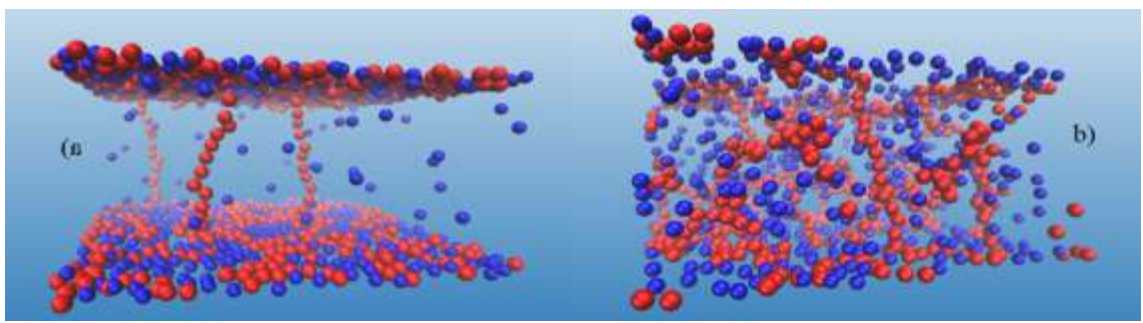

*Figure 10. Snapshots of the simulations with 300 polyelectrolyte beads; on the left (Fig 10a) there are 3 chains of P100 on Au, while on right side (Fig 10b) there are 3 chains of P100 polyelectrolytes on $SiO_2$. The P100 polyelectrolytes form bridges, which join both surfaces; this phenomenon produces major counter ion adsorption on the $SiO_2$ wall. These bridges were found in all simulations with the P100 polyelectrolyte.*

As shown in *Fig 10*, there appears the formation of well defined layers (*Fig 10a*) of polyelectrolytes, counterions and "bridges" between the surfaces, while in *Fig 10b* the adsorption is on the SiO$_2$ surface, but shows bridging also. These bridges appear as a consequence of the confinement because the degree of polymerization of the polyelectrolytes is ten times larger than the length of the simulation box along the *z* - axis. While the Au surface promotes the formation of well-defined monolayers, the SiO$_2$ surface presents diffuse layers. The difference between these two systems lies in the surface charge defined in the simulation; the SiO$_2$ surface has ten times more charge than the Au surface which, as discussed above, limits the polyelectrolyte adsorption. Parts of these long chains, and their counterions, are then located around the center of simulation box, as shown in Fig. 2b, forming the bridges. Although parts of the short chains, and their counterions, are also observed in Fig. 2a, the bridges are not present because of the short size of the polymers. The polyelectrolyte size has then a dominant effect in these simulations. This result anticipates the formation of bridges among chains when two particles with long adsorbed polymers are close.

**IV. Conclusions**

In this work we have shown that the adsorption of polyelectrolytes of the P10 or P100 types of polyelectrolytes on a given surface is very similar. However, comparing adsorption on the Au and on the SiO$_2$ surfaces leads to different results. The Au surface attracts more the polyelectrolytes and develops a well defined monolayer, while adsorption on the SiO$_2$ surface is not as strong as on Au. The leading effect of the polyelectrolyte's polymerization degree is the formation of bridges that reveal the role of the counterions when two surfaces with long adsorbed polyelectrolytes are close. This study can be important to help understand the interaction between nano colloidal particles, and may also be of interest to researchers studying the assembly of polymers on surfaces [32, 33], such as in ring formation [34].

## Acknowledgments

MABA thanks PRODEP DSA/103.5/15/3894 and CA - INGENIERÍA DE PROCESOS QUÍMICOS Y AMBIENTALES - UASLP for a postdoctoral position. RC thanks E. Mayoral for valuable discussions. J. Limón (IF UASLP) is acknowledged for computational support. The authors are especially indebted to F. Alarcón for very helpful input and advice. Thanks are also due to UNISON´s Ocotillo Cluster for valuable computer time.